\documentclass[12pt,twoside]{article}
\usepackage{inputenc}
\usepackage[english,russian]{babel} 
\usepackage{vestnik_eng}
\usepackage{amsfonts}
\usepackage{amsmath}
\usepackage{amssymb}
\usepackage{amsthm}
\usepackage{graphicx}          
\usepackage[usenames]{color}
\begin{document}
\def\No{\textnumero}
\def\Re{\mathop{\rm Re}\,}
\def\Im{\mathop{\rm Im}\,}
\def\dom{\mathop{\rm dom}\,}
\def\dist{\mathop{\rm dist}}
\def\grad{\mathop{\rm grad}}
\renewcommand{\proof}{\vspace{2mm}\hspace{-7mm}\textit{Proof.}}
\renewcommand{\endproof}{\begin{flushright} \vspace{-2mm}$\Box$\vspace{-4mm}
\end{flushright}}
\newcommand{\phan}{\hspace*{0cm}}
\newcommand{\comment}{}
\begin{flushleft}
\textbf{MSC 82D15, 76A20, 76M25} 
\end{flushleft}
\author{E.O. Agafonnikova, {\rm  Volgograd State University, Volgograd, Russian Federation, agafonnikova@volsu.ru}, \\ A.Yu. Klikunova, {\rm  Volgograd State University, Volgograd, Russian Federation, klikunova@volsu.ru} \\ A.V. Khoperskov, {\rm  Volgograd State University, Volgograd, Russian Federation, khoperskov@volsu.ru}}
\title{A COMPUTER SIMULATION OF THE VOLGA RIVER HYDROLOGICAL REGIME: A PROBLEM OF WATER-RETAINING DAM OPTIMAL LOCATION}
\maketitle{A COMPUTER SIMULATION OF THE VOLGA RIVER HYDROLOGICAL REGIME: A PROBLEM OF WATER-RETAINING DAM OPTIMAL LOCATION}
\begin{abstract} \begin{tabular}{p{0mm}p{139mm}}
&\noindent {\footnotesize \qquad We investigate of a special dam optimal location at the Volga river in area of the Akhtuba left sleeve beginning (7 \, km to the south of the Volga Hydroelectric Power Station dam). We claim that a new water-retaining dam can resolve the key problem of the Volga-Akhtuba floodplain related to insufficient water amount during the spring flooding due to the overregulation of the Lower Volga.
By using a numerical integration of Saint-Vacant equations we study the water dynamics across the northern part of the Volga-Akhtuba floodplain with taking into account its actual topography.
As the result we found an amount of water $V_A$ passing to the Akhtuba during spring period for a given water flow through the Volga Hydroelectric Power Station (so-called hydrograph which characterises the water flow per unit of time).
By varying the location of the water-retaining dam $ x_d, y_d $ we obtained various values of $V_A (x_d, y_d) $ as well as various flow spatial structure on the territory during the flood period.
Gradient descent method provide us the dam coordinated with the maximum value of ${V_A}$.
Such approach to the dam location choice let us to find the best solution, that the value $V_A$ increases by a factor of 2.
Our analysis demonstrate a good potential of the numerical simulations in the field of hydraulic works.

\qquad\keywordsanglish{hydrodynamic simulation, Saint-Venant equations, numerical model, optimization, hydrology.}}
\end{tabular}\end{abstract}

\markboth{E.O. Agafonnikova, A.Yu. Klikunova, A.V. Khoperskov}{Computer simulation of the Volga River}


\section*{Introduction}
\hspace{0.7 cm}The unique landscape of 20000\,km$^2$ Volga-Akhtuba floodplain (VAF) depends on special features of the interfluve hydrological regime.
 During the spring flood period the area between the Volga and Akhtuba rivers is heavily flooded \cite{Khrapov2013ame, Middelkoop-etal-2015VAP, Ladjel-2016}, that ensures a special composition of flora and fauna and possibility of agricultural use of the area including the development of magnificent gardens and melon fields.
The floodplain {is also} the basis for the fish reproduction at the Lower Volga region \cite{Gorski-etal-2012VAP}.
Nowadays the overregulation of the Volga-Kama basin hydrological regime by 22 Hydroelectric Power Plants leads to {the} VAP degradation.

Various approaches to the problem solution have beed proposed.
Let us point out the attempts to construct of so-called optimal hydrograph $Q(t)$ \cite{Voronin-etal2012PU, Chen-etal-2015Optimal, Haghighi2015optim} which is close to the natural and ensures preservation of the natural rate.
Despite the progress in the construction of the mathematical models and models of hydrological regime control and the territory as a whole \cite{Borsche-Klar-2014, vvu2016VAP, Izem-etal-2016num, Bulatov-Elizarova-2016SW}, there is a great difficulty of their practical implementation due to the conflict of various agents aspiration (energetics, environmental protection organizations, fish industry, agriculture, inhabitants, safety of reservoirs and etc.).
 {Over the last} decades the situation {becomes even} more complicated due to changes in the Volga riverbed below the dam {mainly} because of the violation of the spring flood natural process.
 In this study we discuss the possibility of the floodplain hydrological regime improvement by the construction of the water-retaining dam at the Volga riverbed close to the beginning of the Akhtuba left sleeve {which is located at approximately} 7\,km below the Volga Hydroelectric Power Station dam
 (Fig.\,1).
The main aim of the work is construction of a mathematical model for the estimation of the dam optimal location {which provide} the larger water flow rate in the Akhtuba during the spring flood.

\begin{figure}
\centering\includegraphics[width=0.9\hsize]{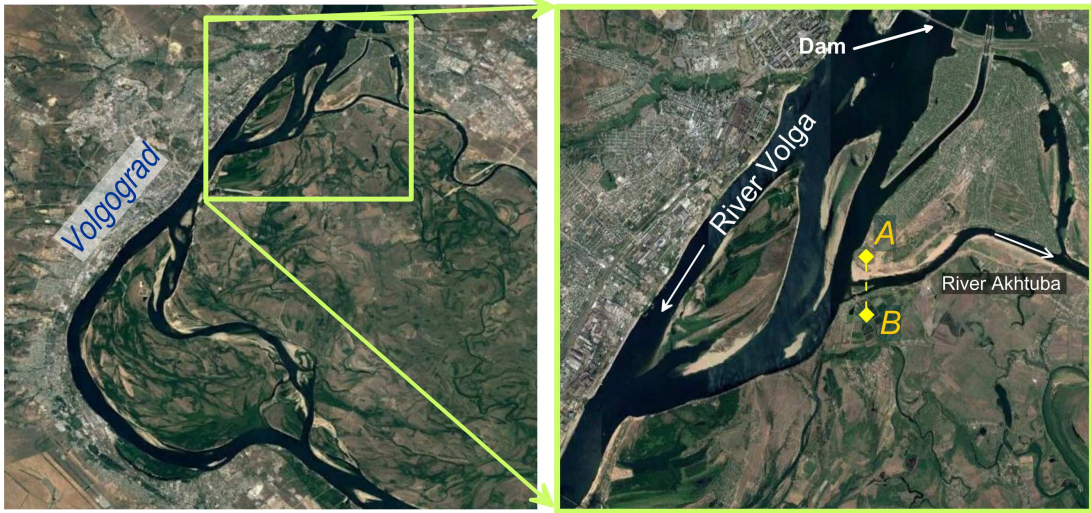}\vskip 0.05\hsize
\caption{The map of the northern part of the Volga-Akhtuba floodplain with the topography.
The insert shows the part of the Volga with the Akhtuba springhead.
}
\label{fig-AKK-commap}
\end{figure}

\section{The hydrodynamic model}
\hspace{0.7 cm}At first, we briefly describe the hydrodynamic model underlying in the base of our research.
We use the Saint-Venant equations for the shallow water dynamics at a given topography $b(x,y)$ \cite{Khrapov2013ame, vvu2016VAP, DKhKh2017ccis}:
\begin{equation}\label{eq-Sen-Venan-H}
    \frac{\partial H}{\partial t} + \frac{\partial uH}{\partial x} + \frac{\partial vH}{\partial y} = \sigma(x,y,t) \,,
\end{equation}
\begin{equation}\label{eq-Sen-Venan-u}
    \frac{\partial (uH)}{\partial t} + \frac{\partial (u^2H)}{\partial x} + \frac{\partial (uvH)}{\partial y} = - gH\frac{\partial (H+b)}{\partial x} + 2vH\Omega_E \sin(\theta) + f_x^{\rm{fric}} +f_{\sigma x} \,,
\end{equation}
\begin{equation}\label{eq-Sen-Venan-v}
    \frac{\partial (vH)}{\partial t} + \frac{\partial (uvH)}{\partial x} + \frac{(v^2 H)}{\partial y} = - gH\frac{\partial (H+b)}{\partial y} -2uH\Omega_E\sin(\theta) + f_y^{\rm{fric}} +f_{\sigma y}\,,
\end{equation}
where $H$ is the water depth, $u$ and $v$~are the velocity of $x$- and $y$-components (averaged vertically), $\sigma$ is the source function, $g $~is referred to the gravitational acceleration, $\Omega_E$ is the  Earth's angular velocity, $\theta$ is the latitude, the values of $f_{\sigma x}$ and $ f_{\sigma y}$ describe the water impulse associated with the sources $\sigma$.
For the bottom friction force vector we use the Chezy's model \cite{Khrapov2013ame}:
\begin{equation}\label{eq-ffric}
 f_x^{\rm{fric}} = - \frac{u}{2}\sqrt{u^2+v^2}\, H\Lambda\,, \quad
 f_y^{\rm{fric}} = - \frac{v}{2}\sqrt{u^2+v^2}\, H\Lambda,
\end{equation}
with the value of hydraulic friction $\Lambda = 2gn_M^2/H^{4/3}$ and the Manning roughness coefficient~$n_M$.

Value of $ n_M $ is determined by the surface properties and generally it depends on the coordinates.
 Moreover, for the unsteady flow regime the Manning coefficient can vary with time \cite{Khrapov2013ame}.
The hydrograph through the Volga Hydroelectric Power Station dam allows to set the value of $\sigma=\frac{dQ_0}{dS}$ (where $ dS = dx \, dy $ is an elementary area).
Under the VAP conditions the value of $ n_M $ for the Volga riverbed vary in the range $0.02 - 0.07$~\cite{Khrapov2013ame}.

For the numerical solution of equations (\ref{eq-Sen-Venan-H})--(\ref{eq-Sen-Venan-v}) we apply our combined Lagrangian-Eulerian method (cSPH-TVD) \cite{Khrapov-etal-2011csphtvd} which uses the benefits of Smoothed Particle Hydrodynamics at different time steps and Total Variation Diminishing (see the detailed description in Refs.~\cite{Khrapov2013ame, Khrapov-etal2016schem}).

The most important positive characteristics of the cSPH-TVD approach are the following:

\noindent --- an adequate calculation of the dynamic boundaries between wet and dry beds in case of non-stationary fluxes through the strongly inhomogeneous bottom (even through the discontinuous topography);

\noindent --- calculation for subcritical (with Froude number $Fr =\sqrt{u^2+v^2}/\sqrt{gH}<1$) and supercritical ($Fr>1$) fluxes without isolation of these zones;

\noindent
---	numerical scheme of CSPH-TVD is conservative, well-balanced and has the second order of accuracy for smooth solutions {the first order accuracy approximation} in the vicinity of breaks and fracture profiles.

Both non steady solutions and strong heterogeneity of the topography require the specific boundary conditions formulation.
 Ref.~\cite{DKhKh2016udm} is dedicated to the application of the boundary conditions for the same aims by using the conditions of  ``waterfall'' type, {which we adopted} in the current study.

The software implementation by using parallel technologies on graphics accelerators is presented in ref.~\cite{DKhKh2017ccis}.
All basic calculations were performed on the GPU NVIDIA Tesla K80 \cite{DKhKh2017ccis}.
We use Digital Elevation Models (DEM) with spatial resolution $ \Delta{x} = \Delta{y} = 50$\,m and 25\,m, {which is based on combination of} several geodata: ASTER GDEM 2 (Global DIGITAL Elevation Model), SRTM X-SAR (Shuttle Radar Topography Mission) and Sentinel-1 SAR data, topographic data for coastlines of the hydrological system, our GPS / GLONASS measurements.
 To improve the model we use the sailing directions and special numerical hydrodynamical simulations allowing us to compare our results with observational data.

\section{The optimal dam location}
\begin{figure}[th]
\centering{\includegraphics[width=0.8\hsize]{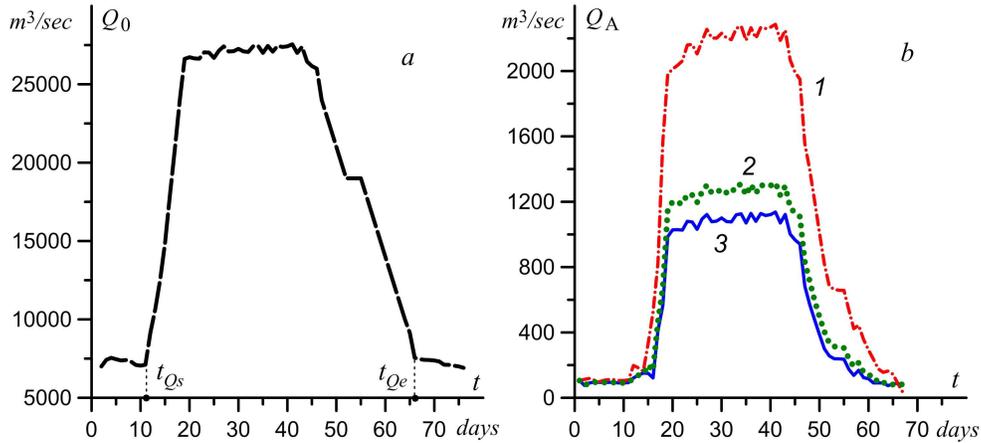}}
 \vskip 0.05\hsize
\caption{
The spring hydrograph $Q_0(t)$ of  2016 through the Volga Hydroelectric Power Station dam {adopted} in the model (\textit{a}). The hydrographs of Akhtuba $Q_A$ at different dam positions (\textit{b}): \textit{1} --- the best dam position; \textit{2} --- without dam; \textit{3} --- the worst dam position.}
\label{fig-AKK-hydrograph}
\end{figure}

\begin{figure}[th]
\centering\includegraphics[width=1.0\hsize]{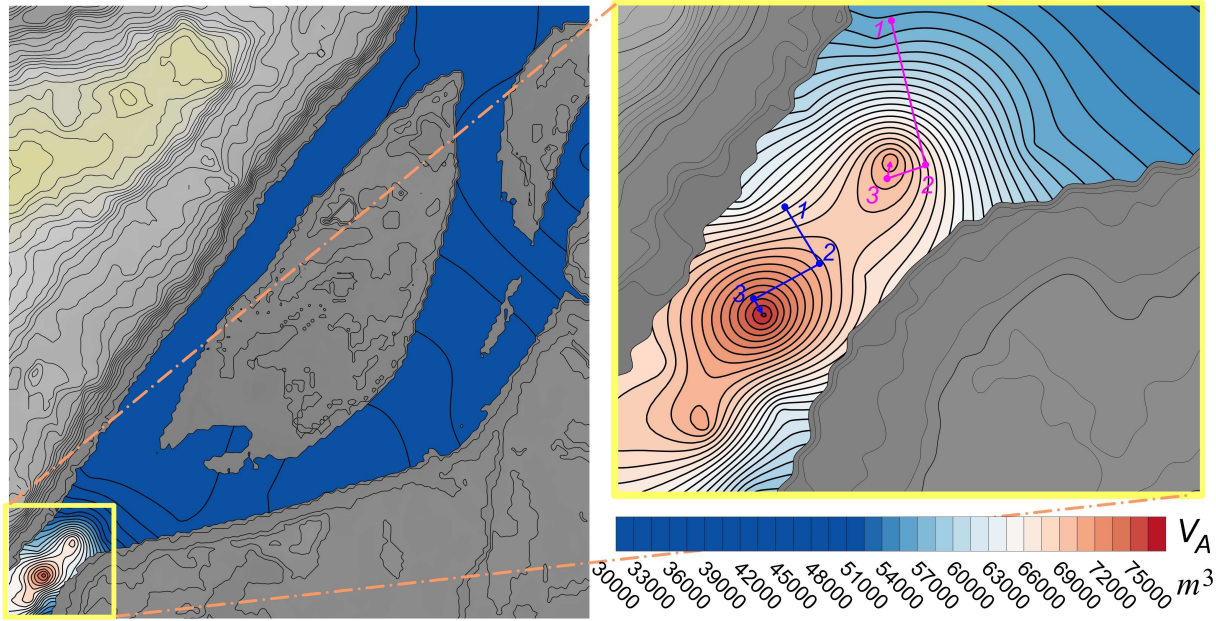}\vskip 0.05\hsize
\caption{
The results of numerical hydrodynamical simulations for the function $V_A(x_D, y_D)$ and examples of the iterative procedure for the calculation of $V_A^{(\max)}$ in case of $L_d=300$\,m.}
\label{fig-AKK-mapQxy}
\end{figure}

\hspace{0.7 cm}For a given Volga Hydroelectric Power Station hydrograph $Q_0(t)$ we can calculate the water volume entering the Akhtuba during the spring flood (see Figure\,\ref{fig-AKK-commap}):
\begin{equation}\label{eq-VAchtuba}
    V_A = \int\limits_{A}^{B} \int\limits_{t_{Qs}}^{t_{Qe}} H(x,y,t) \left( u\cdot n_x + v\cdot n_y \right) \,dt\,d\ell \,,
\end{equation}
 where the unit vector $\vec{n} = (n_x, n_y)$ is a normal to the Akhtuba river section line $(A,B)$ (see. Fig.\,\ref{fig-AKK-mapQxy}\,b), $ t_{Qs} $ and $ t_{Qe}$~are the water release at beginning and final time, respectively (see. Fig.\,\ref{fig-AKK-hydrograph}).
 {Functions} $H(x, y, t)$, $u(x, y, t)$ and $v(x, y, t)$ are calculated by using the hydrodynamical model (\ref{eq-Sen-Venan-H})\,--\,(\ref{eq-Sen-Venan-v}).
We set up the dam of length $L_d$ at the Volga riverbed close to the Akhtuba springhead at the point  $ (x_d, y_d) $ (dam's center) which is perpendicular to the coastlines. The dam {affects on the} flow structure and on the value of $V_A$.
 For a given $L_d$ and $Q_0(t)$ we have function $V_A(x_d, y_d)$.
We calculate the water-retaining dam optimal location for the specific riverbed area $S_A$ according to the following condition:
\begin{equation}\label{eq-maxVA}
    V_A^{(\max)} = \max\limits_{(x_d,y_d)\in S_A} V_A(x_d,y_d) \,,
\end{equation}
$S_A$~--- is part of the Volga riverbed near the beginning of Akhtuba (between about 7\,km downstream and 5\,km upstream).

In figure 3 we show the $V_A^{(\max)}$ search procedure by using the gradient descent method:
\begin{equation}\label{eq-grad-method}
 \vec{r}_d^{\,(k+1)} = \vec{r}_d^{\,(k)} + \lambda^{(k)} \, {\rm{grad}}(V_A(\vec{r}_d^{\,(k)})) \,.
\end{equation}
Finite difference approximation for the gradient calculation {is used in the following form}:
\begin{equation}\label{eq-grad-numer}
 {\rm{grad}}(V_A(\vec{r}_d)) \simeq \left\{ \frac{V_A(x_d+\delta{x},y_d)-V_A(x_d,y_d)}{\delta{x}} ; \frac{V_A(x_d,y_d+\delta{y})-V_A(x_d,y_d)}{\delta y} \right\}
\end{equation}
on the meshgrid $x_{i+1} = x_i + \Delta{x}$, $y_{j+1} = y_j + \Delta{y}$.
{Our test numerical simulations demonstrated} that the relations $\delta{x}=\delta{y}=2\Delta{x}=2\Delta{y}$ are reasonable choice.

The symbols in Figure~\ref{fig-AKK-mapQxy}\,{\textit{a}} indicate the positions of the iterative procedure (\ref{eq-grad-method}), (\ref{eq-grad-numer}).
The choice of $\lambda_k$ parameter and the convergence of calculations are achieved by using the steepest descent method.

\section{Results and discussion}
\hspace{0.7 cm}By means of numerical simulations we have studied the problem of the water-retaining dam location optimization at the Volga riverbed with the aim to increase the water amount at the left sleeve of river Akhtuba. We summarize our results as following.
There are several dam locations providing the appearance of the water volume local maximum $V_A$ in the Akhtuba during the spring flood.
The positions $ (x_{dm}^{(\max)}, y_{dm}^{(\max)})$ ($m = 1,2,3$) are located about 6 km downstream from the Akhtuba's beginning, that caused by the Volga riverbed structure at the area due to the large Volga width nearby the Akhtuba and large island (see Fig.\,1).
The best solution is $V_A^{(\max)} = 75000$\,m$^3$.
As a result, we {can achieve} a factor of 1.6 for maximum value of the hydrograph $Q_A(t)$ (see Figure 2) and almost factor of 2 for $V_A$ from $V_{A0}=43000$\,m$^3$ up to $V_{A0}=75000$\,m$^3$ in comparison to the absence of the extra dam ($V_{A0} = 43000$\,m$^3$).

For a fixed set of free parameters any deviation of the dam orientation from the perpendicular relatively to the coastlines reduces the value $V_A$.
The optimum location slightly depends on the dam size.
It should be noted that there are some dam positions {that determine} the value of $V_A$ {which is smaller than in the case} of absence of the water-retaining dam.
Our analysis is robust to small scale perturbations of the digital topography.
A small variation of the function $b(x,y)$ conserve the approximate optimal solution $ (x_{d}^{(\max)}, y_{d}^{(\max)}) $, {but the issue} still requires further investigation.

\vspace{5mm}{\bf Acknowledgements.} {\it   We are thankful to the Ministry of Education
and Science of the Russian Federation (project 2.852.2017/4.6). The study was supported by the Supercomputing Center of Lomonosov Moscow State University. EOA is thankful to the RFBR (grants 16-07-01037, 15-45-02655).}

\begin{biblio_lat}

\bibitem{Khrapov2013ame}  Khrapov S., Pisarev A., Kobelev I., Zhumaliev A., Agafonnikova E., Losev A., Khoperskov A.  The Numerical Simulation of Shallow Water: Estimation of the Roughness Coefficient on the Flood Stage \textit{Advances in Mechanical Engineering}, 2013, vol.\,5, Article ID 787016 pp.\,1--11. DOI: 10.1155/2013/787016\vspace{-2mm}\vspace{-2mm}

\bibitem{Middelkoop-etal-2015VAP} Middelkoop H., Alabyan A.M., Babich D.B., Ivanov V.V.  Post-dam Channel and Floodplain Adjustment along the Lower Volga River, Russia. \textit{Geomorphic Approaches to Integrated Floodplain Management of Lowland Fluvial Systems in North America and Europe}. Springer, 2015, pp.\,245--264. DOI: 10.1007/978-1-4939-2380-9\_10\vspace{-2mm}

\bibitem{Ladjel-2016} Ladjel M. Lamination method of flood wadis and projection of the
laminated flood hydrograph. \textit{Journal of Fundamental and Applied Sciences}, 2016, vol.\,8, no.\,1, pp.\,83--91.  DOI: 10.4314/jfas.v8i1.6\vspace{-2mm}

\bibitem{Gorski-etal-2012VAP} Gorski K., van den Bosch L.V., van de Wolfshaar K.E., Middelkoop H., Nagelkerke L.A.J., Filippov O.V. et al.
 Post-damming flow regime development in a large lowland river (Volga, Russian Federation): implications for floodplain inundation
and fisheries. \textit{River Research and Applications},
2012, vol.\,28, no.\,8, pp.\,1121--1134. DOI: 10.1002/rra.1499.\vspace{-2mm}

\bibitem{Voronin-etal2012PU}  Voronin A.A., Eliseeva M.V., Khrapov S.S., Pisarev A.V., Khoperskov A.V. [The Regimen Control Task in The Eco-Economic System ``Volzhskaya Hydroelectric Power Station --- the Volga-Akhtuba Floodplain''. II. Synthesis of Control System]. Problemy Upravleniya [Control Sciences], 2012, no.6, pp.\,19--25. (in Russian)\vspace{-2mm}

\bibitem{Chen-etal-2015Optimal} Chen D., Li R., Chen Q., Cai D. Deriving Optimal Daily Reservoir Operation Scheme with Consideration of Downstream Ecological Hydrograph Through A Time-Nested Approach. \textit{Water Resources Management}, 2015, vol.\,29, no.9, pp.\,3371--3386. DOI 10.1007/s11269-015-1005-z\vspace{-2mm}

\bibitem{Haghighi2015optim} Haghighi A.T., Kl{\o}ve B. Development of monthly optimal flow regimes for allocated
environmental flow considering natural flow regimes and several
surface water protection targets. \textit{Ecological Engineering}, 2015, vol.\,82, pp.\,390-399. DOI: 10.1016/j.ecoleng.2015.05.035\vspace{-2mm}

 \bibitem{Borsche-Klar-2014}Borsche R., Klar A. Flooding in urban drainage systems: coupling hyperbolic conservation laws for sewer systems and surface flow. \textit{International Journal for Numerical Methods in Fluids}, 2014, vol.\,76, pp.\,789--810. DOI: 10.1002/fld.3957\vspace{-2mm}

\bibitem{vvu2016VAP}  Voronin, A.A., Vasilchenko, A.A., Pisarev, A.V., Khrapov, S.S., Radchenko, Yu.E. [Designing Mechanisms of the Hydrological Regime Management of the Volga-Akhtuba Floodplain Based on Geoinformation and Hydrodynamic Modeling].
     \textit{Vestnik Volgogradskogo gosudarstvennogo universiteta. Seriya 1. Matematika. Fizika}. [Science Journal of VolSU. Mathematics. Physics], 2016, no.\,1 (32) pp.\,24--37. DOI: 10.15688/jvolsu1.2016.1.3 (in Russian)\vspace{-2mm}

\bibitem{Izem-etal-2016num} Izem N., Seaid M., Wakrim M. A discontinuous Galerkin method for two-layer shallow water equations. \textit{Mathematics and Computers in Simulation}, 2016, vol.\,120, pp.\,12–23. DOI: 10.1016/j.matcom.2015.04.009\vspace{-2mm}

\bibitem{Bulatov-Elizarova-2016SW} Bulatov O.V., Elizarova T.G. Regularized shallow water equations for numerical simulation of flows with a moving shoreline. \textit{Computational Mathematics and Mathematical Physics}, 2016, vol.\,56, no.\,4, pp.\,661--679. DOI: 10.1134/S0965542516040047\vspace{-2mm}

\bibitem{DKhKh2017ccis}  D'yakonova, T.A., Khoperskov A.V., Khrapov S.S. Numerical model of shallow water: the use of GPUs NVIDIA CUDA \textit{Communications in Computer and Information Science}, 2017, vol.\,687, pp.\,132--145\vspace{-2mm}

\bibitem{Khrapov-etal-2011csphtvd} Khrapov S.S., Khoperskov A.V., Kuz'min N.M., Pisarev A.V., Kobelev I.A. [A numerical scheme for simulating the dynamics of surface water on the basis of the combined SPH-TVD approach]. \textit{Vychislitel'nye Metody i Programmirovanie}. [Numerical Methods and Programming], 2011, vol.\,12, pp.\,282--297 (in Russian)\vspace{-2mm}

\bibitem{Khrapov-etal2016schem}
 Khrapov S.S., Kuzmin N.M., Butenko M.A. [The Comparison of Accuracy and Convergence for the CSPH--TVD Method and Some Eulerian Schemes for Solving Gas-Dynamic Equations]. \textit{Vestnik Volgogradskogo gosudarstvennogo universiteta. Seriya 1. Matematika. Fizika.} [Science Journal of VolSU. Mathematics. Physics], 2016, no.\,6 (37), pp.\,166--173. DOI: 10.15688/jvolsu1.2016.6.15 (in Russian)\vspace{-2mm}

\bibitem{DKhKh2016udm}  D'yakonova, T.A., Khrapov S.S., Khoperskov A.V. [The problem of boundary conditions for the shallow water equations]. \textit{Vestnik Udmurtskogo Universiteta. Matematika. Mekhanika. Komp'yuternye Nauki} [The Bulletin of Udmurt University. Mathematics. Mechanics. Computer Science], 2016, vol.\,26, no.\,3, pp.\,401--417. DOI: 10.20537/vm160309 (in Russian)\vspace{-2mm}

\end{biblio_lat}

\end{document}